\documentclass[preprint,10pt,5p]{elsarticle}

\usepackage{amssymb}

\usepackage{lineno}
\usepackage{indentfirst}
\usepackage{amsmath}
\usepackage{algorithm}
\usepackage[noend]{algpseudocode}
\usepackage{multicol}
\usepackage{float}
\usepackage{graphicx}
\usepackage{comment}

\usepackage{subcaption}

\usepackage{hhline}

\usepackage{hyperref}

\usepackage{booktabs}
\usepackage{mathcomp}
\usepackage{textcomp}

\journal{}

\begin{document}

\begin{frontmatter}

\title{\textit{Aedes aegypti} Egg Counting with Neural Networks for Object Detection}

\address[label1]{Dom Bosco Catholic University, Campo Grande, Brazil}
\address[label2]{Federal University of Mato Grosso do Sul, Campo Grande, Brazil}

\author[label1]{Micheli Nayara de Oliveira Vicente}
\ead{ra162169@ucdb.br}

\author[label1]{Gabriel Toshio Hirokawa Higa\corref{cor1}}
\ead{ra190503@ucdb.br}
\cortext[cor1]{Corresponding author}

\author[label1]{João Vitor de Andrade Porto}
\ead{ra688593@ucdb.br}

\author[label2]{Higor Henrique Picoli Nucci}
\ead{higor.nucci@ufms.com}

\author[label1]{Asser Botelho Santana}
\ead{ra171911@ucdb.br}

\author[label1]{Karla Rejane de Andrade Porto}
\ead{karla.porto2@docente.suafaculdade.com.br}

\author[label1]{Antonia Railda Roel}
\ead{arroel@ucdb.br}

\author[label1,label2]{Hemerson Pistori}
\ead{pistori@ucdb.br}

\begin{abstract}
\textit{Aedes aegypti} is still one of the main concerns when it comes to disease vectors. Among the many ways to deal with it, there are important protocols that make use of egg numbers in ovitraps to calculate indices, such as the LIRAa and the Breteau Index, which can provide information on predictable outbursts and epidemics. Also, there are many research lines that require egg numbers, specially when mass production of mosquitoes is needed. Egg counting is a laborious and error-prone task that can be automated via computer vision-based techniques, specially deep learning-based counting with object detection. In this work, we propose a new dataset comprising field and laboratory eggs, along with test results of three neural networks applied to the task: Faster R-CNN, Side-Aware Boundary Localization and FoveaBox.
\end{abstract}


\end{frontmatter}


\section{Introduction}
\label{intro}

\textit{Aedes} (\textit{Stegomyia}) \textit{aegypti} (Linnaeus, 1762) (Diptera: Culicidae) is an insect associated with the infestation and transmission of several diseases, including dengue, chikungunya fever and the Zika virus. In Brazil, disease outbursts and epidemics related to the \textit{A. aegypti} result in high expenses to the national health system. The problems caused by them can be considered nationwide and chronic. According to Siqueira Junior \textit{et al.}~\cite{siqueira_junior2022epidemiology}, in the last years, Brazil went through four epidemics with over one million cases of dengue, in the years of 2013, 2015, 2016 and 2019, and overall costs of the disease ranged from US\$ 516.79 million in 2009 to US\$ 1,688.3 million in 2013\footnote{Adjusted to 2019 USD, according to the authors.}.

Among the several control strategies implemented and used in protocols by the Vector Control Center of Campo Grande, state of Mato Grosso do Sul, Brazil, that follows the works of Garcia \textit{et al.}~\cite{Garcia2019}, the methods believed to have the highest efficacy are ones that use indices such as the Larval Index Rapid Assay for \textit{Aedes aegypti} (LIRAa) and the Breteau Index, which require counting the number of eggs, usually in ovitraps. The use of these and similar indices to predict disease outbursts is a successful strategy, whose importance was emphasized, for instance, in the study of Sanchez-Gendriz \textit{et al.}~\cite{sanchez_gendriz2022data_driven}, who thoroughly studied the possibility of using time sequence data for prediction in a Brazilian city. When egg numbers are used, the counting is usually done manually with the assistance of a magnifier or a microscope.

Counting \textit{A. aegypti} eggs is also useful in scientific researches other than those aiming directly at predicting outbursts. Some researches have also used eggs of other species of the \textit{Aedes} genus. Bakran-Lebl \textit{et al.}~\cite{bakran_lebl2022first}, for example, counted 63.287 mosquito eggs in a research on invasive \textit{Aedes} species in Austria, and that of Brisco \textit{et al.}~\cite{brisco2023field}, who counted \textit{Aedes} eggs within a research to assess the results of a vector control policy in Hawaii, albeit in a much smaller scale.

The idea of automating the laborious task of counting eggs of \textit{A. aegypti} is by no means a new one. As discussed by Brun \textit{et al.}~\cite{brun2020revisao}, researches on the application of classic computer vision and machine learning techniques to the task appeared as early as 2008. On the other hand, according to the review, the use of deep learning techniques is more recent, going back to 2019. Most of the works focus on eggs that are laid by female mosquitoes in areas where outbreaks are likely. 

A work that used deep learning was that by de Santana \textit{et al.}~\cite{de2019solution}, who provided a realistic dataset and tested the performance of some algorithms to count eggs of \textit{A. aegypti}. Another work was that by Garcia \textit{et al.}~\cite{Garcia2019}, who used images of ovitraps as a way to measure egg deposition. The researchers used a strategy of segmenting and classifying, and report that over 90\% of the eggs were found. Furthermore, they indicated three images in which the counting was much worse. In these, they argue that the main difficulties are the presence of dirtiness and the high density of eggs. The images presented to support the claim do show high countings of eggs. However, they do not seem to be tightly clustered, as the one presented by us. Furthermore, although they counted 90\% of the eggs (with an IoU threshold of 0.3, leading to low precision results), their methodology is lacking in robustness, since the experiment was conducted without repetitions and the test set comprised only 30 images.

Currently, efforts to increase the performance of systems to recognize eggs are still ongoing, for instance in the work of Gumiran \textit{et al.}~\cite{gumiran2022aedes}, who investigated the visual features of eggs. The most important visual features, according to the authors, are: shape, size and color. In the current stage, there are also researches that aim at designing more practical systems. For instance, Abad-Salinas \textit{et al.}~\cite{salinas2022computer} presented a prototype of an intelligent ovitrap that uses a Raspberry Pi for counting eggs. Another example is the work of Javed \textit{et al}.~\cite{javed2023eggcountai}, where a software for counting \textit{Aedes} eggs was proposed. The authors separated two groups of images, one of which they considered micro images, with up to 215 eggs. The other group was that of macro images, with up to 3658 eggs per image. The authors report an overall accuracy of 98.8\% for micro images, and 96.06\% for macro images. However, from the presented images one can observe that the eggs were not as tightly clustered as they are in our case. Also, since the test set contains only 10 images and no repetitions were done, the methodology is not robust, and the results, albeit very high, were not properly validated.

Given the state-of-the-art, it is noticeable that automatically counting eggs laid in laboratory conditions is a task that has not yet been properly addressed. The importance of counting eggs obtained in field notwithstanding, there are research lines that require counting eggs laid in laboratory, mainly for testing diverse techniques. For instance, Iyyappan \textit{et al.}~\cite{iyyappan2022oviposition} used egg numbers to evaluate the effectiveness of different organic infusions used to attract female mosquitoes. Khan \textit{et al.}~\cite{khan2023assessment} carried out a study in which the egg number was used to compare the attractiveness of different colors and materials on mosquitoes. In this last case, the experiments were separately executed in laboratory conditions and in field.

It is well acknowledged in the literature that the task of counting eggs laid in the field, when carried out as described above, is a laborious, physically demanding, slow and error-prone one. For field conditions, it is clear by now that computational techniques are a viable solution for egg counting~\cite{brun2020revisao}, and there are no \textit{a priori} reasons to assume that they do not work for laboratory conditions. There are reasons, however, to assume that the tasks are not trivially interchangeable, since the difference between conditions leads to differences in the visual aspect of the context that surrounds the eggs, and also of the eggs themselves.

\begin{figure}[!h]
\begin{center}

\includegraphics[height=0.4\linewidth,width=0.4\linewidth]{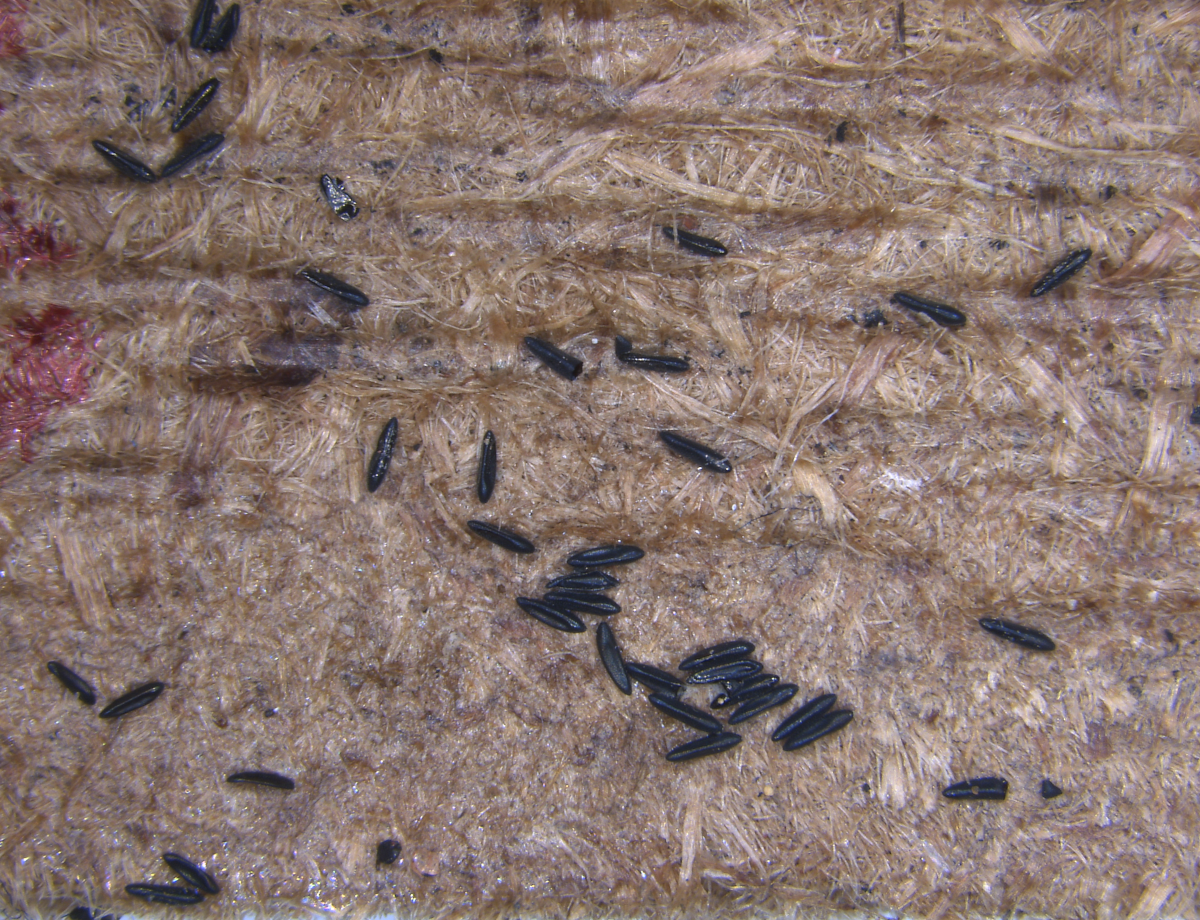}
\includegraphics[height=0.4\linewidth,width=0.4\linewidth]{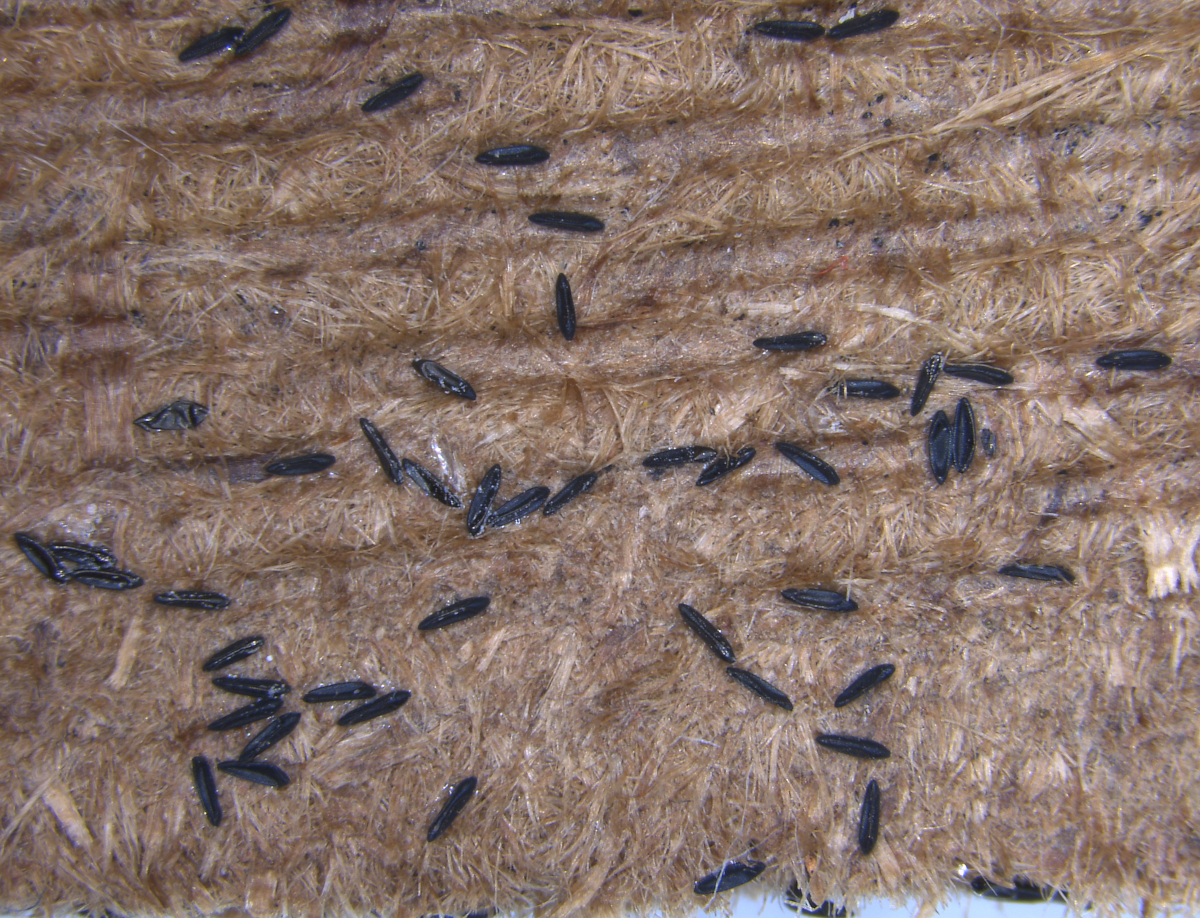}
\includegraphics[height=0.4\linewidth,width=0.4\linewidth]{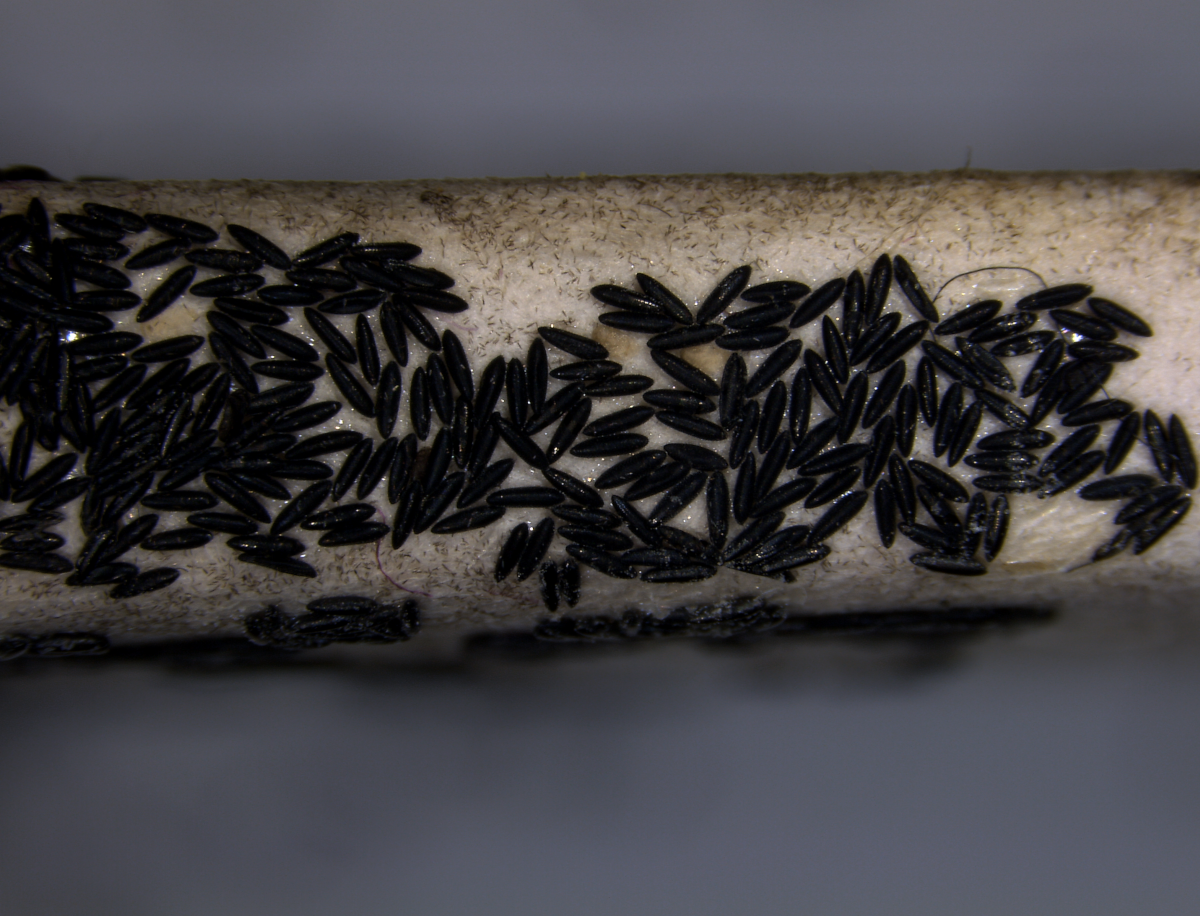}
\includegraphics[height=0.4\linewidth,width=0.4\linewidth]{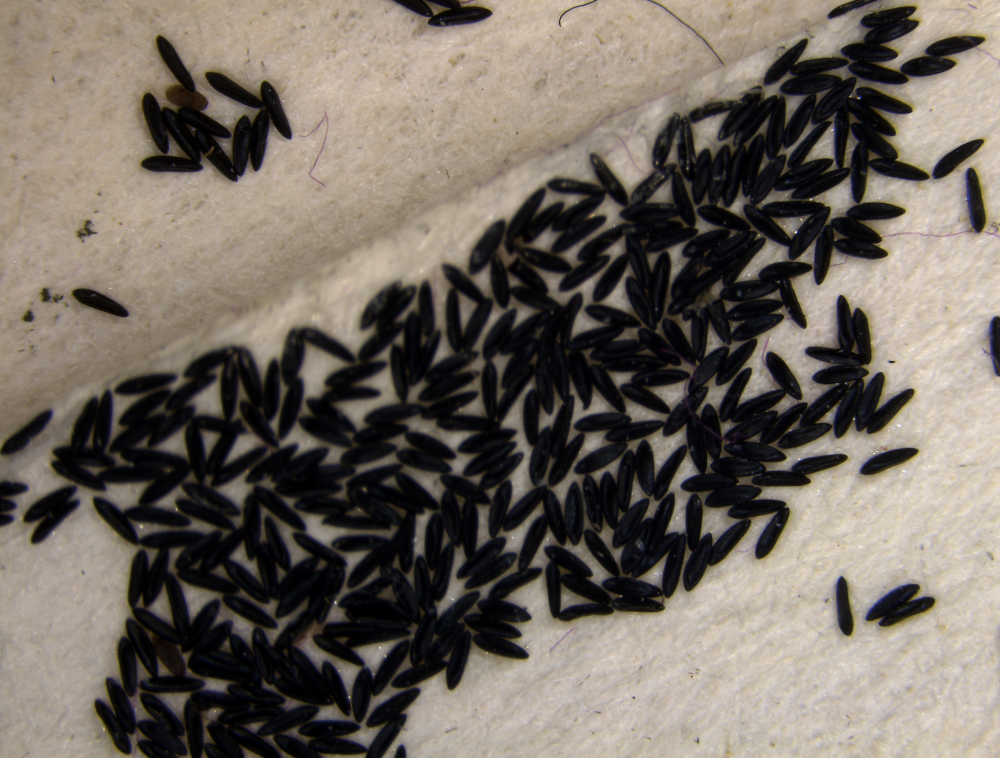}

\end{center}
\caption{Collected eggs. Images were taken with the Leica MC170 HD stereomicroscope. The images in the first row show eggs collected in the field. The ones in the second row are laboratory eggs.}
\label{fig:eggs_raw}
\end{figure}

That being said, in this work, we present a new image dataset for \textit{A. aegypti} egg counting, which is non-trivially different from those already available in that it comprises both situations: eggs collected in the field and eggs mass-produced in a laboratory. The most visible, and crucial, difference between the corresponding images is the quantity of eggs. Figure~\ref{fig:eggs_raw} shows samples of images of both situations. The difference in quantity can be easily seen, clusters of eggs are a common situation in laboratory conditions, given the necessity of mass production for different kinds of tests. Other differences are the presence of dirt and the physical condition of the eggs.

Among the machine learning techniques, those that involve image processing, belonging to the field of computer vision, can be considered the most adequate ones, given the nature of the task, which is doable through object detection techniques. The state-of-the-art in object detection techniques is achieved through deep learning techniques. Therefore, we also present the results achieved by three neural networks applied to the task: Faster R-CNN, SABL and FoveaBox. 

\section{Materials and methods}
\label{materials_methods}

\subsection{The image dataset}

Initially, the eggs of \textit{A. aegypti} were collected in field, in Campo Grande, MS, Brazil, by agents of the Center of Epidemiologic Control of Vectors of the Municipal Health Secretariat (CCEV/SESAU). The eggs were collected by using ovitraps, with pallets that were partially submerged in hay water to facilitate egg laying. 

Following the work of Ricci \textit{et al.}~\cite{anaricci}, the eggs were left to mature for seven days, protected from light and humidity. For maturation, a BOD incubator was used, with a temperature of $27 \pm 2$ \textcelsius, RH of $75 \pm 5\%$ and photophase of 12 hours. Then, they were separated between viable and non-viable. From the viable eggs, 200 were positioned on the bottom of a plastic recipient containing 1.5L of dechlorinated water of pH between 6.5 and 7. Then, 0.002g of ration per larva was added to the recipient. The hatched larvae were taken care of until the formation of pupae, which were collected and transferred to cages until the adult stage of the mosquitoes. The male mosquitoes were fed a sugar solution (8\% of sucrose) and the females received bird blood meal for 40 minutes in intervals of two days.

From the eggs collected in field, the eggs of the first generation (F1) were obtained in laboratory. The eggs of F1 were also collected with ovitraps, and were matured following the same protocol utilized for the field eggs, the only difference being that filter paper was used in the ovitraps, to keep the adequate humidity levels and to facilitate hatching. In the adult stage, the female mosquitoes of F1 were also fed a grass solution after the blood meal to increase egg-laying with greater viability.

\begin{figure}[t]
\begin{center}
    \includegraphics[height=0.4\linewidth,width=0.4\linewidth]{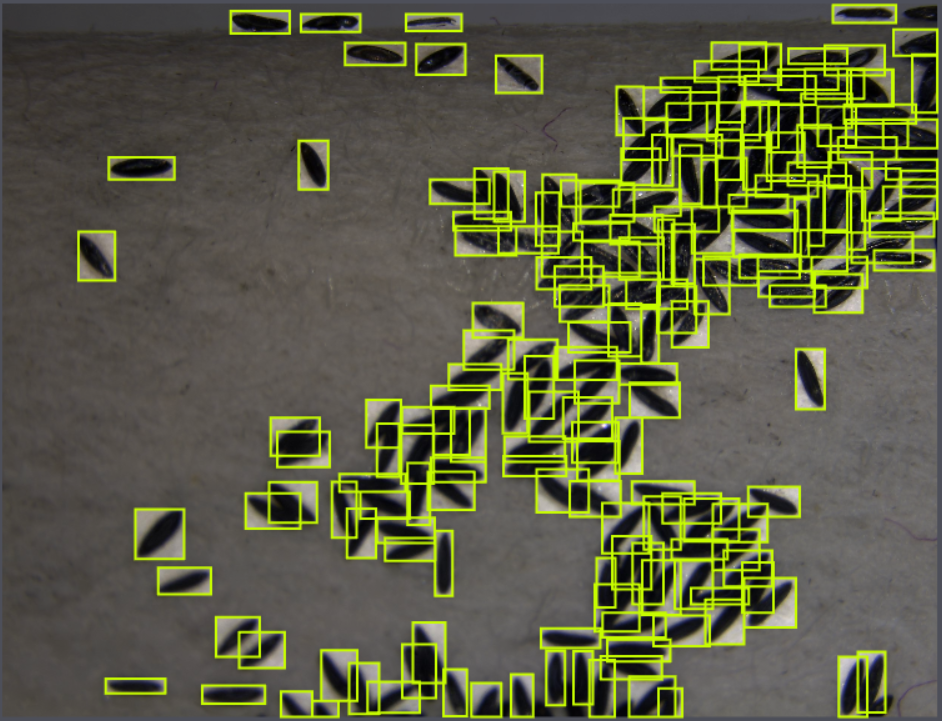}
    \includegraphics[height=0.4\linewidth,width=0.4\linewidth]{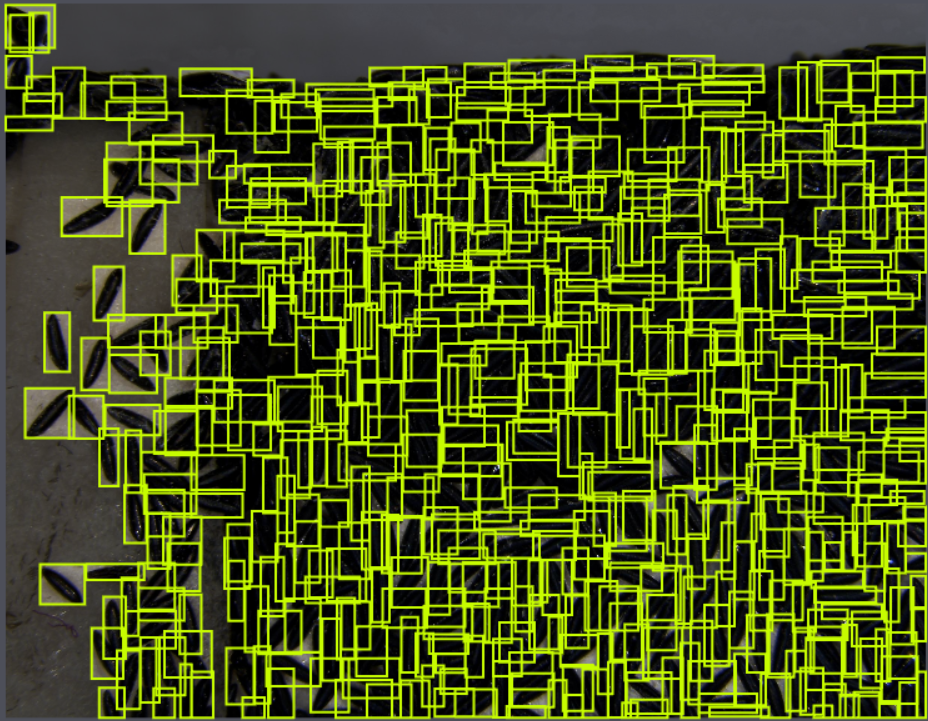}
\end{center}
\caption{Two sample images annotated with Roboflow.}
\label{fig:ex_anotada}
\end{figure}

The pictures of each set of eggs were taken before the eggs were left to mature. The images were made with a Leica MC170 HD stereomicroscope in the Laboratory of Entomology (B09) of the Dom Bosco Catholic University. Figure~\ref{fig:eggs_raw} shows examples of the images thereby obtained. The images were then annotated for object detection with \textit{Roboflow}\footnote{Available here: \url{https://roboflow.com/}}. Samples of annotated images can be seen in Figure~\ref{fig:ex_anotada}. The annotated images were exported in COCO-JSON format. The image dataset contains 247 images. Of these, 123 are of field eggs and 124 are of F1. In total, there are $12.513$ annotated \textit{A. aegypti} eggs.

\subsection{Neural networks}
\label{neural_nets}
In the field of computer vision, counting tasks have been approached through object detection techniques that make use of convolutional neural networks (CNN). In this work, we test the performance of three architectures commonly utilized for object detection: Faster R-CNN, Side-Aware Boundary Localization (SABL) and FoveaBox.

The Faster R-CNN, proposed by Ren \textit{et al.}~\cite{article_faster}, can be considered the third version of the Region-Based Convolutional Neural Network (R-CNN)~\cite{article_rcnn}. It was proposed after the Fast R-CNN, which is the second version of the R-CNN~\cite{article_fast_rcnn}. As the name indicates, the main objective of the different versions was to improve computation speed. To do that, the Faster R-CNN introduced the usage of a neural network for region proposal, making convolutional layers shareable between the region proposal network (RPN) and the Fast R-CNN module. In this work, it is used with a ResNet50-FPN backbone\footnote{This is implemented in MMDetection as faster\_rcnn\_r50\_fpn\_1x\_coco, available here: \url{https://github.com/open-mmlab/mmdetection/tree/master/configs/faster_rcnn}}.

The Side-Aware Boundary Localization (SABL) was proposed by Wang \textit{et al.}~\cite{article_sabl}. In itself, SABL is a new way of refining the bounding box localization, having been presented as an alternative to the usual bounding box regression. The authors use the notion of buckets, divisions on each side of the map of the region of interest, predicting first the bucket to which a boundary of the box belongs, and then refining the prediction with respect to the bucket. In this work, we use RetinaNet with SABL and ResNet50-FPN as backbone\footnote{Implemented in MMDetection as sabl\_retinanet\_r50\_fpn\_1x\_coco, available here: \url{https://github.com/open-mmlab/mmdetection/tree/master/configs/sabl}}.

Finally, the third object detection network tested in this work is FoveaBox. FoveaBox was proposed by Kong \textit{et al.}~\cite{kong2020foveabox} and belongs to the category of detection networks that do not use anchors. It was inspired by the fovea of the human eyes, the basic idea being to predict the center of an object in the image, if it exists, along with two points defining the bounding box. In this work, the tested version uses a ResNet50-FPN as backbone\footnote{Implemented in MMDetection as fovea\_r50\_fpn\_4x4\_1x\_coco, available here: \url{https://github.com/open-mmlab/mmdetection/tree/master/configs/foveabox}}.

The first architecture, Faster R-CNN was chosen as a classic and accomplished detection network. It is used to compare the performance of SABL and FoveaBox, which are more recent\footnote{Both were originally proposed in 2020, while the Faster R-CNN was already being used in 2015.} and proposed new techniques for object localization, which may have an impact on the network performance when applied to the task of egg counting.

\subsection{Experimental Setup}
\label{experimental_setup}

To evaluate the neural networks listed in Section~\ref{neural_nets}, we used the implementations available in the MMDetection package. The hyperparameters were left as is, including the image size, set to (1333, 800), which is usual for testing in object detection tasks. The only hyperparameter that was changed is the maximum number of possible detections during test, which was set to 1000. This is 10 times the default value in the implementation. Limits for training were not changed, since they are larger by default. This was done in order to make the test fit to images with a large number of annotated eggs, which go beyond 500 in some images of the dataset. 

All the neural networks were optimized with Stochastic Gradient Descent (SGD). For the Faster R-CNN, a learning rate of 0.02 was used. For both SABL and FoveaBox, a learning rate of 0.01 was used. These values were chosen because they were used in the original articles of the architectures. However, differently from the original works, we did not use learning rate scheduling. The other hyperparameters of the SGD optimizer were also kept as the default values: momentum as 0.9 and weight decay as 0.0001 for all architectures. These choices were also taken because searching for optimal hyperparameters goes beyond the scope of this work. 

The architectures were tested through a stratified 10-fold cross validation strategy. The training was performed in 30 epochs. In each epoch, 20\% of the training images were used for validation. To evaluate the architectures, the following metrics were calculated on the test sets after each run: mAP50, mAP75, mAP, MAE, RMSE, precision, recall and f-score, as well as Pearson's coefficient of correlation (r). Although Pearson's r is not a measurement of error \textit{per se} (and it is possible for a set of predictions to present high error with high positive correlation), it was included in this study due to its straightforwardness: if the neural network counts eggs adequately, any variation in the number of eggs in an image must imply a variation in the number of counted eggs in the exact same proportion. Ideally, the correlation between groundtruths and predictions should assume the greatest possible value (and the error equal zero).

After testing, boxplots were also generated. An ANOVA hypothesis testing was used to evaluate the architectures, with a chosen threshold of 5\%. As the task at hand is object counting, MAE, RMSE and Pearson's r were taken as dependent variables for the ANOVA, which was independently applied for each of the metrics. Tukey's Honestly Significant Difference (TukeyHSD) was used as a post hoc test when ANOVA results were significant. Other metrics were further evaluated when the discussion thus required. After the cross validation, the counting was also evaluated as one group, apart from the division in folds. The MAE, RMSE and Pearson'r were calculated for them, and scatter plots of groundtruth and predictions were generated, along with the best fit line.

An in-depth analysis was then conducted on the results of the most promising architecture (understood as that which achieved the smallest average RMSE). The objective of this analysis was to identify and summarize the difficulties involved in the task (thus the choice for the most promising architecture). For this in depth-analysis, we separated the results according to the number of eggs in the image: first, into groups of images with up to 100 eggs, images with over 100 eggs up to 300 eggs, and images with over 300 eggs; then, we separated the images with up to 100 eggs into a group with up to 50 eggs and another one with more than 50 eggs (up to 100). This procedure was taken in order to better evaluate how the increase in the number of eggs influences the performance of the networks (and through this, what are the difficulties posed by them). The separation into these groups was a procedural choice taken so that a more focused analysis would be possible.

\section{Results and discussion}
\label{results_discussion}

\begin{table}[t]
\caption{Statistics for MAE, RMSE and Pearson's coefficient of correlation used to evalute the performance of the neural networks in counting tasks. These values were calculated within the 10-fold cross validation strategy.}
\label{table:count_metrics}
\centering
\resizebox{9cm}{!}{
\begin{tabular}{|c|c|c|c|c|}
\toprule

\multicolumn{5}{c}{\textbf{MAE}} \\ \hline
\textbf{Architecture} & \textbf{Median} & \textbf{IQR} & \textbf{Mean} & \textbf{SD} \\ \hline \hline
\textbf{Faster R-CNN} & 8.958 & 12.667 & 12.171 & 7.741 \\ \hline
\textbf{SABL} & 11.146 & 13.419 & 14.201 & 8.632 \\ \hline
\textbf{FoveaBox} & \textbf{6.854} & \textbf{8.116} & \textbf{9.213} & \textbf{5.347} \\ \hline

\multicolumn{5}{c}{\textbf{RMSE}} \\ \hline
\textbf{Architecture} & \textbf{Median} & \textbf{IQR} & \textbf{Mean} & \textbf{SD} \\ \hline \hline
\textbf{Faster R-CNN} & 28.678 & 32.200 & 34.684 & 23.307 \\ \hline
\textbf{SABL} & 36.301 & 30.052 & 40.195 & 24.022 \\ \hline
\textbf{FoveaBox} & \textbf{19.725} & \textbf{20.524} & \textbf{23.628} & \textbf{14.587} \\ \hline

\multicolumn{5}{c}{\textbf{Pearson's r}} \\ \hline
\textbf{Architecture} & \textbf{Median} & \textbf{IQR} & \textbf{Mean} & \textbf{SD} \\ \hline \hline
\textbf{Faster R-CNN} & 0.971 & 0.030 & 0.963 & 0.034 \\ \hline
\textbf{SABL} & 0.968 & 0.029 & 0.958 & 0.034 \\ \hline
\textbf{FoveaBox} & \textbf{0.987} & \textbf{0.009} & \textbf{0.989} & \textbf{0.006} \\ \hline
\end{tabular}}

\end{table}

\begin{figure*}[!ht]
    \centering
    \includegraphics[width=1\textwidth]{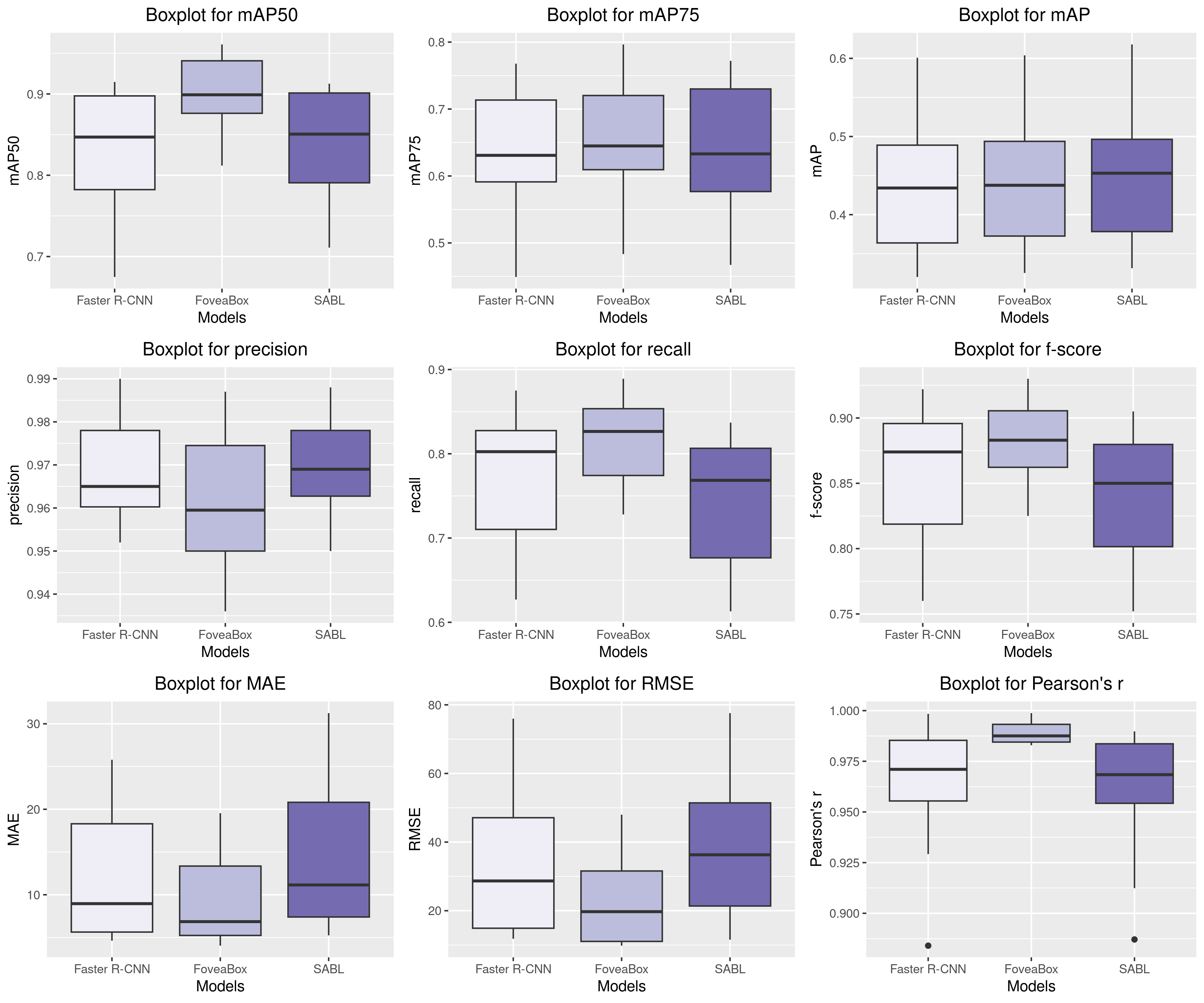}
    \caption{Boxplots for each metric calculated in the experiment.}
    \label{fig:all_boxplots}
\end{figure*}

\begin{figure*}[!th]
    \centering
    \begin{subfigure}[b]{0.3\textwidth}
        \centering
        \includegraphics[width=\textwidth]{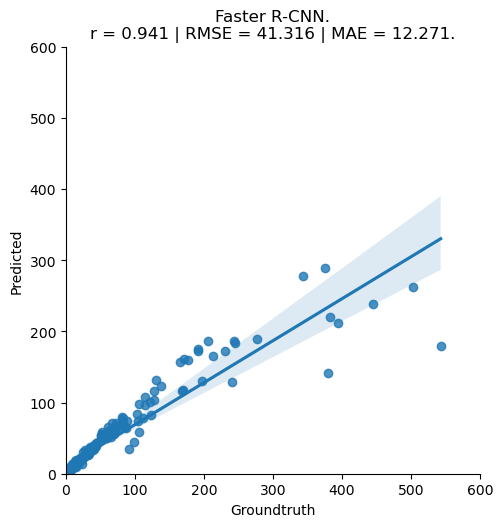}
    \end{subfigure}
    \hfill
    \begin{subfigure}[b]{0.3\textwidth}
        \centering
        \includegraphics[width=\textwidth]{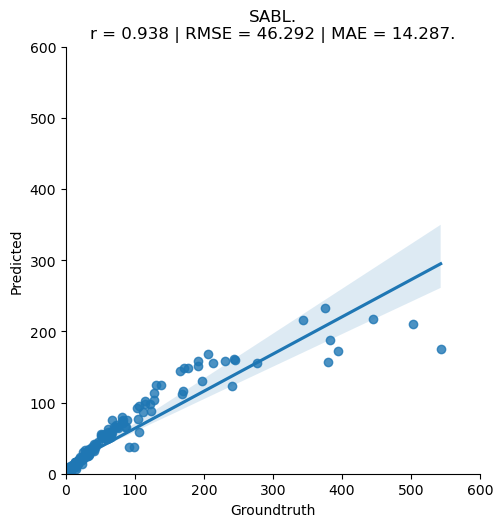}
    \end{subfigure}
    \hfill
    \begin{subfigure}[b]{0.3\textwidth}
        \centering
        \includegraphics[width=\textwidth]{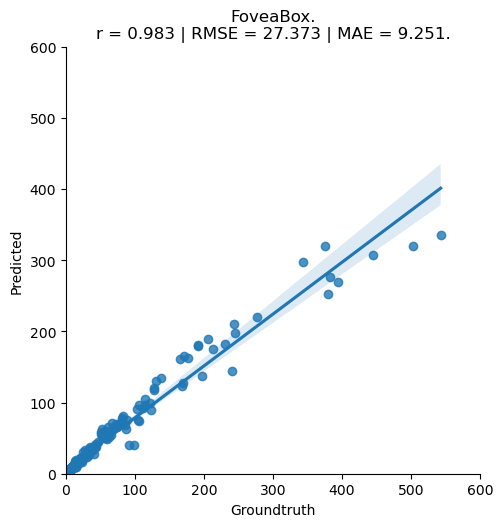}
    \end{subfigure}
    \caption{Scatter plots for each architecture, along with the best fit line. The metrics below the title were calculated differently from those in Table~\ref{table:count_metrics}. Here, they refer to the counting as a whole, not to the results in ten folds.}
    \label{fig:scatter_plots}
\end{figure*}

\begin{figure*}[]
    \centering
    \begin{subfigure}[b]{0.3\textwidth}
        \centering
        \includegraphics[width=\textwidth]{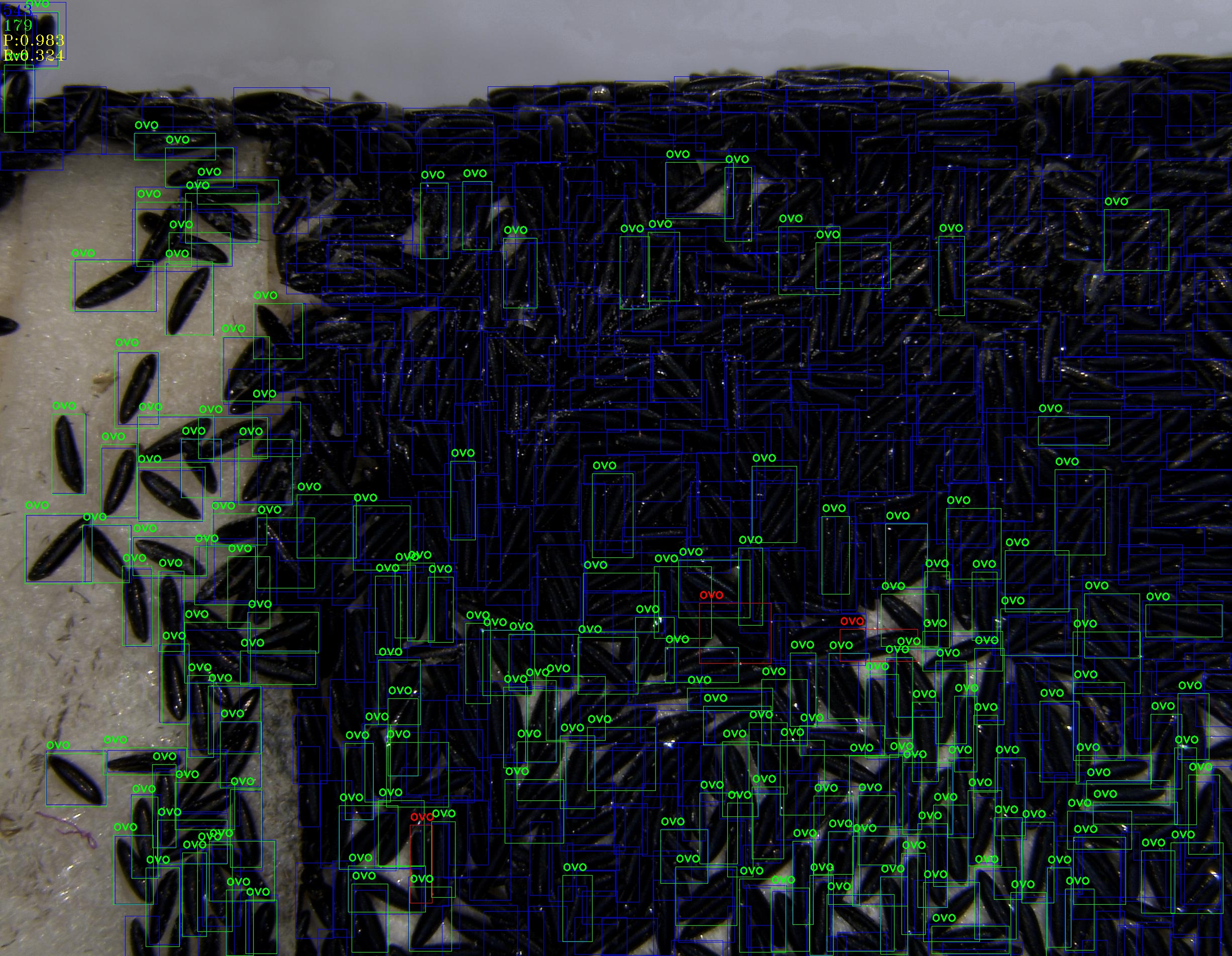}
        \caption{Faster R-CNN}
        \label{fig:faster_543}
    \end{subfigure}
    \hfill
    \begin{subfigure}[b]{0.3\textwidth}
        \centering
        \includegraphics[width=\textwidth]{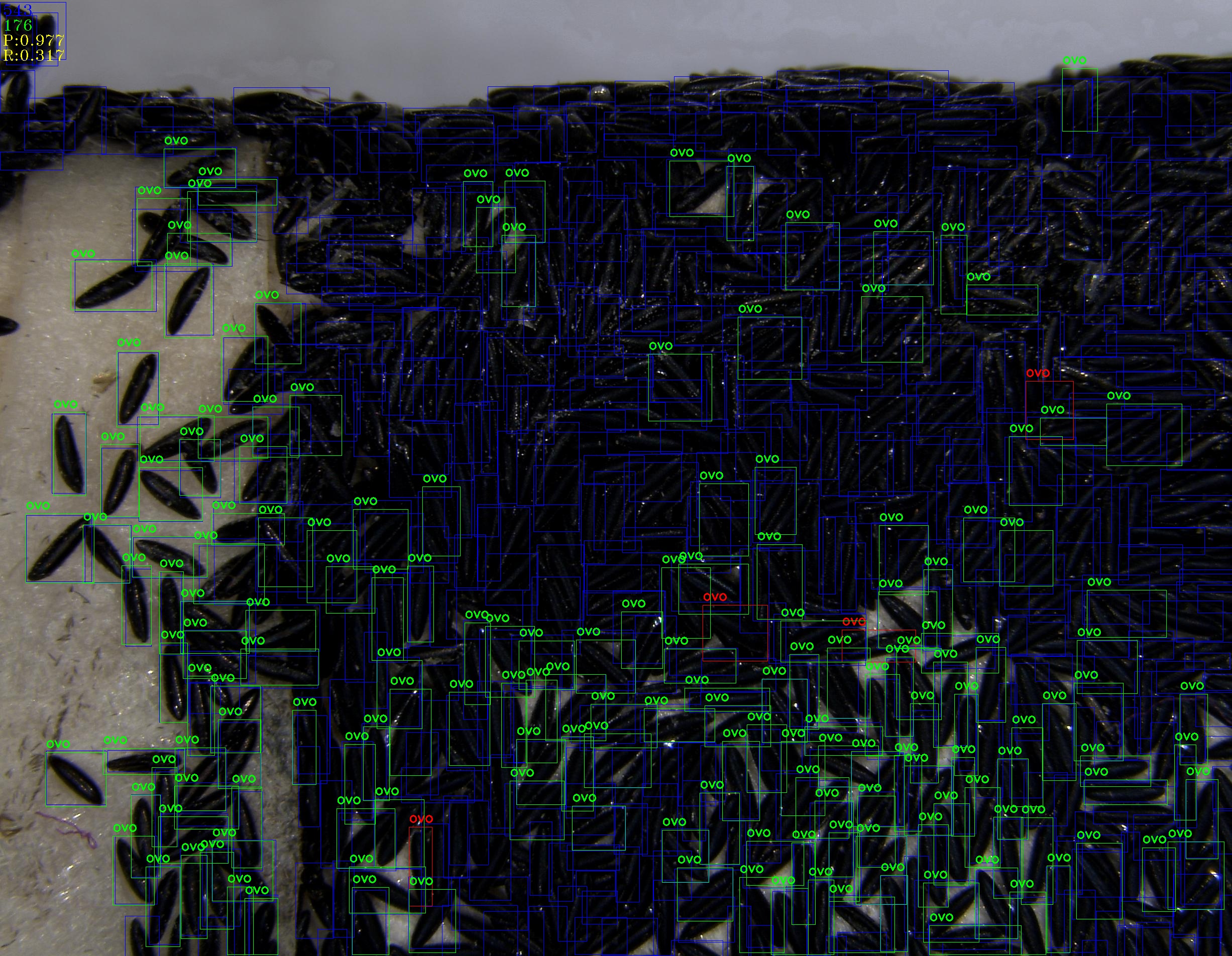}
        \caption{SABL}
        \label{fig:sabl_543}
    \end{subfigure}
    \hfill
    \begin{subfigure}[b]{0.3\textwidth}
        \centering
        \includegraphics[width=\textwidth]{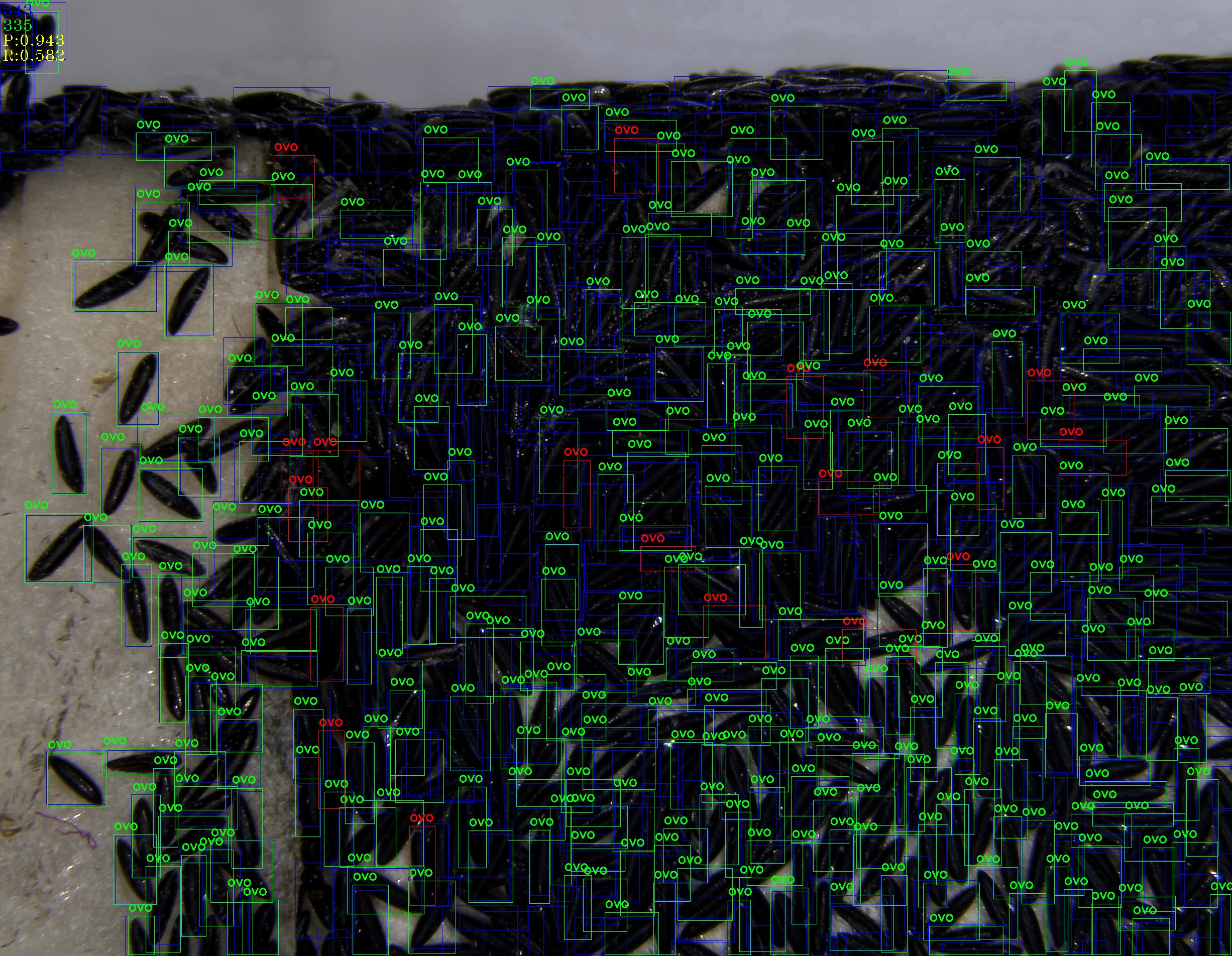}
        \caption{FoveaBox}
        \label{fig:fovea_543}
    \end{subfigure}
    \caption{Annotations (blue), true positives (green) and false positives (red) for each architecture on the image with the highest number of annotations.}
    \label{fig:images_543}
\end{figure*}

Figure~\ref{fig:all_boxplots} shows boxplots for each metric calculated in the experiment across ten runs. Table~\ref{table:count_metrics} shows statistics for the main metrics used to evaluate the networks in the task at hand: MAE, RMSE and Pearson's r. The ANOVA hypothesis test did not indicate difference between the architecture's performances in the case of MAE ($p=0.329$) and RMSE ($p=0.22$). For the coefficient of correlation, the ANOVA yielded a marginally significant result ($p=0.046$), but the TukeyHSD result was actually marginally insignificant when SABL and FoveaBox were compared ($p=0.053$). Furthermore, it was not significant at all when Faster R-CNN was compared both with FoveaBox ($p=0.12$) and with SABL ($p=0.92$).

Figure~\ref{fig:scatter_plots} shows scatter plots of groundtruth vs. prediction for each image in the dataset, for each architecture. The metrics below the title refer to all the images, not to the results per fold (as is the case in Table~\ref{table:count_metrics}). The plots also show the best fit line. As it is also possible to see in Table~\ref{table:count_metrics}, the correlation was, on average, very high. The scatter plots show that the errors tended to be higher in images with more annotations. All in all, FoveaBox achieved a better performance.

The boxplots for Pearson's r in Figure~\ref{fig:all_boxplots} show that SABL and Faster R-CNN had one outlier. Further inspection of the results show that in both cases the outlier r was calculated in fold 6, in which the image with the highest number of eggs (543) was in the test set. In the case of this image, which was taken in laboratory, Faster R-CNN counted 179 eggs (Figure~\ref{fig:faster_543}), SABL counted 176 eggs (Figure~\ref{fig:sabl_543}), and FoveaBox, which did not present an outlier in Figure~\ref{fig:all_boxplots}, managed to count 335 eggs out of 543 annotations (Figure~\ref{fig:fovea_543}). This number may actually be bigger, since inspection of the image shows that some eggs that were considered false positives were, in reality, missed in the labeling process (which is ultimately inevitable, given the quantity of eggs, and reinforces the idea that manually counting eggs is an error-prone task). 

The capacity of FoveaBox of counting more eggs is also shown by its recall results, although the statistical tests were marginally not significant ($p = 0.085$ for ANOVA, with $p = 0.070$ for TukeyHSD when it was compared with SABL and $p = 0.427$ when it was compared with Faster R-CNN), that is, there is no indication that the recall of FoveaBox was better than that of Faster R-CNN, but it arguably was better than that of SABL, marginal significance considered. This can also be seen in the recall boxplots, in Figure~\ref{fig:all_boxplots}. The median of FoveaBox was near the upper quartile of Faster R-CNN, and the IQR was smaller.

\begin{figure}
    \centering
    \includegraphics[width=1\columnwidth]{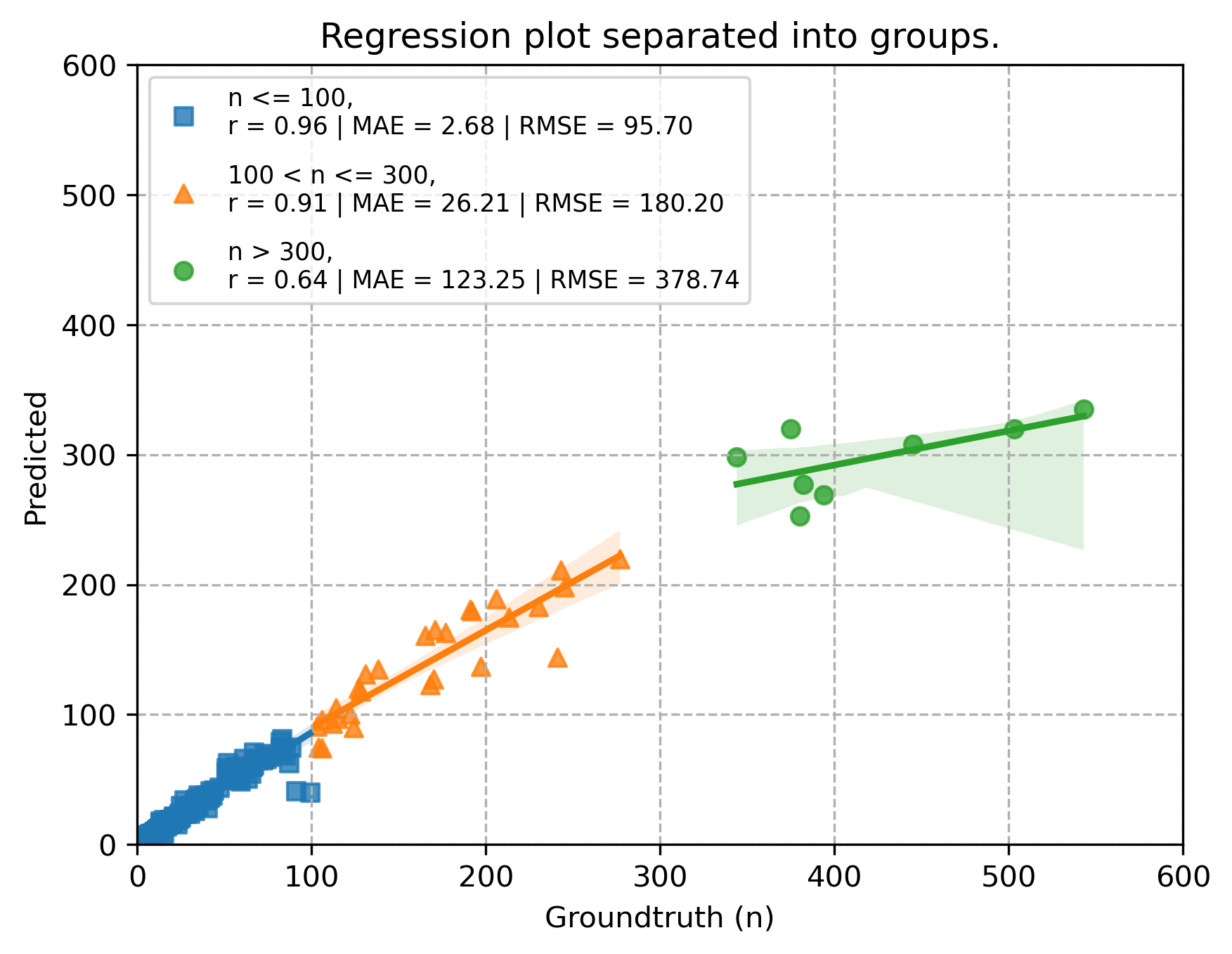}
    \caption{Regression plot for groundtruths and predictions, separated into three groups: the first with images containing up to 100 eggs, the second one with images containing more than 100 eggs up to 300, and the last one with images containing more than 300 eggs. One should notice that the error is much higher for images with more eggs.}
    \label{fig:regplot_all}
\end{figure}

\begin{figure}
    \centering
    \includegraphics[width=1\columnwidth]{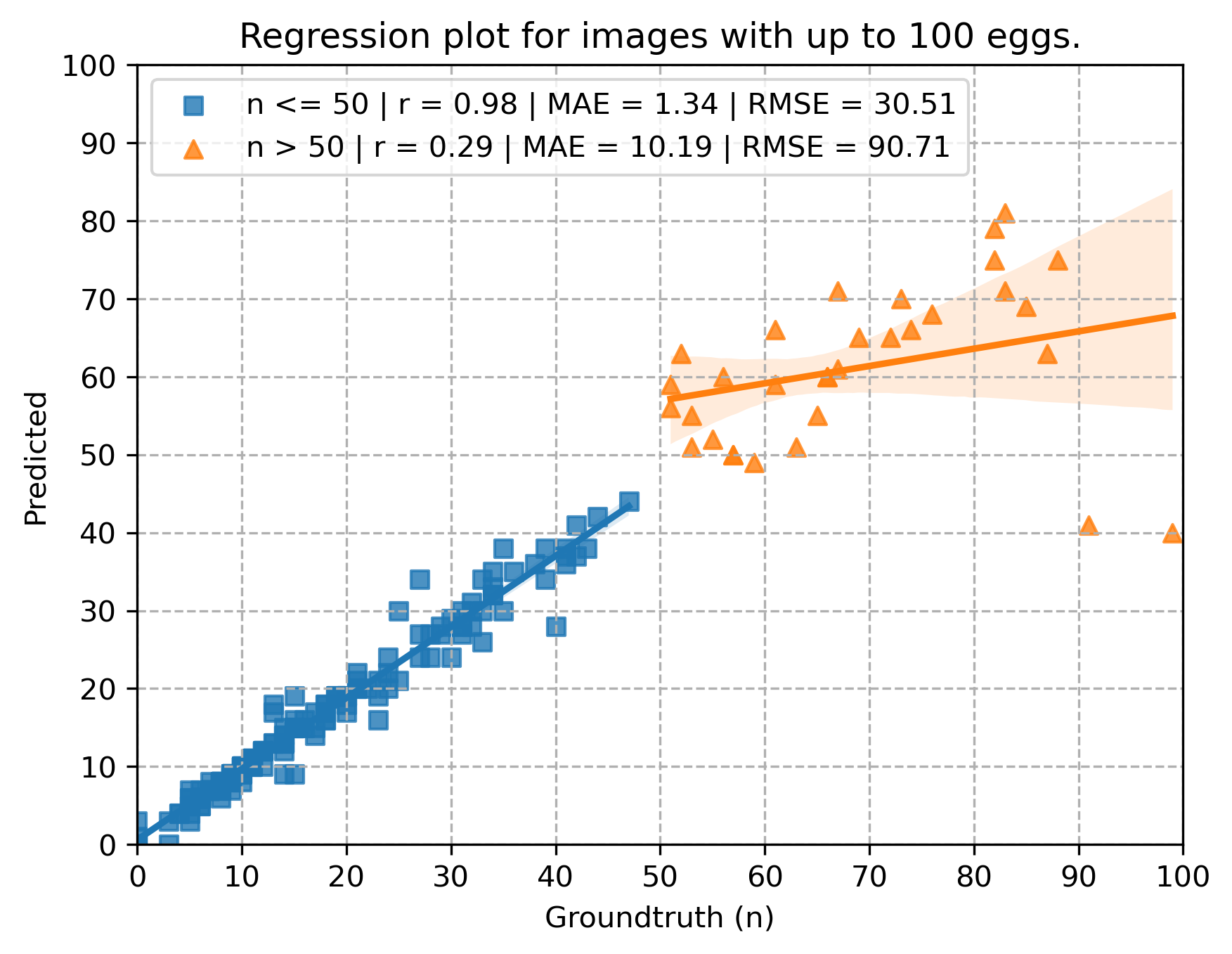}
    \caption{Regression plot for groundtruths and predictions for images containing up to 100 eggs. Again, these were separated into two groups: one for images with up to 50 eggs, and another one for images with more than 50 eggs. In agreement with the last plot, the error is higher for images with more eggs. In this case, Pearson's r was also lower.}
    \label{fig:regplot_up_2_100}
\end{figure}

The results of FoveaBox were selected for an in-depth analysis, since it was found to be the most promising architecture. Figures~\ref{fig:regplot_all} and \ref{fig:regplot_up_2_100} show scatterplots of groundtruths and predictions, along with the corresponding best fit lines for the groups described in Section~\ref{experimental_setup}. One can see from Figure~\ref{fig:regplot_all} that the error is much higher for images with more eggs (almost fourfold for images with more than 300 eggs). When the results for images with less eggs are analysed (in Figure~\ref{fig:regplot_up_2_100}), one can see that it is indeed the images with more than 50 eggs that lead to the worse errors. In this second case, Pearson's r also showed only a weak positive correlation. Nonetheless, an RMSE of over 30 for images with less than 50 eggs can still be considered troublesome (even if an MAE of 1.34 is considered), if the counting is to be used for disease outbreak predictions and scientific research. The situation is even worse for images with more eggs, given the high MAE and RMSE values for images with more eggs.

\begin{figure}
    \centering
    \includegraphics[width=0.7\columnwidth]{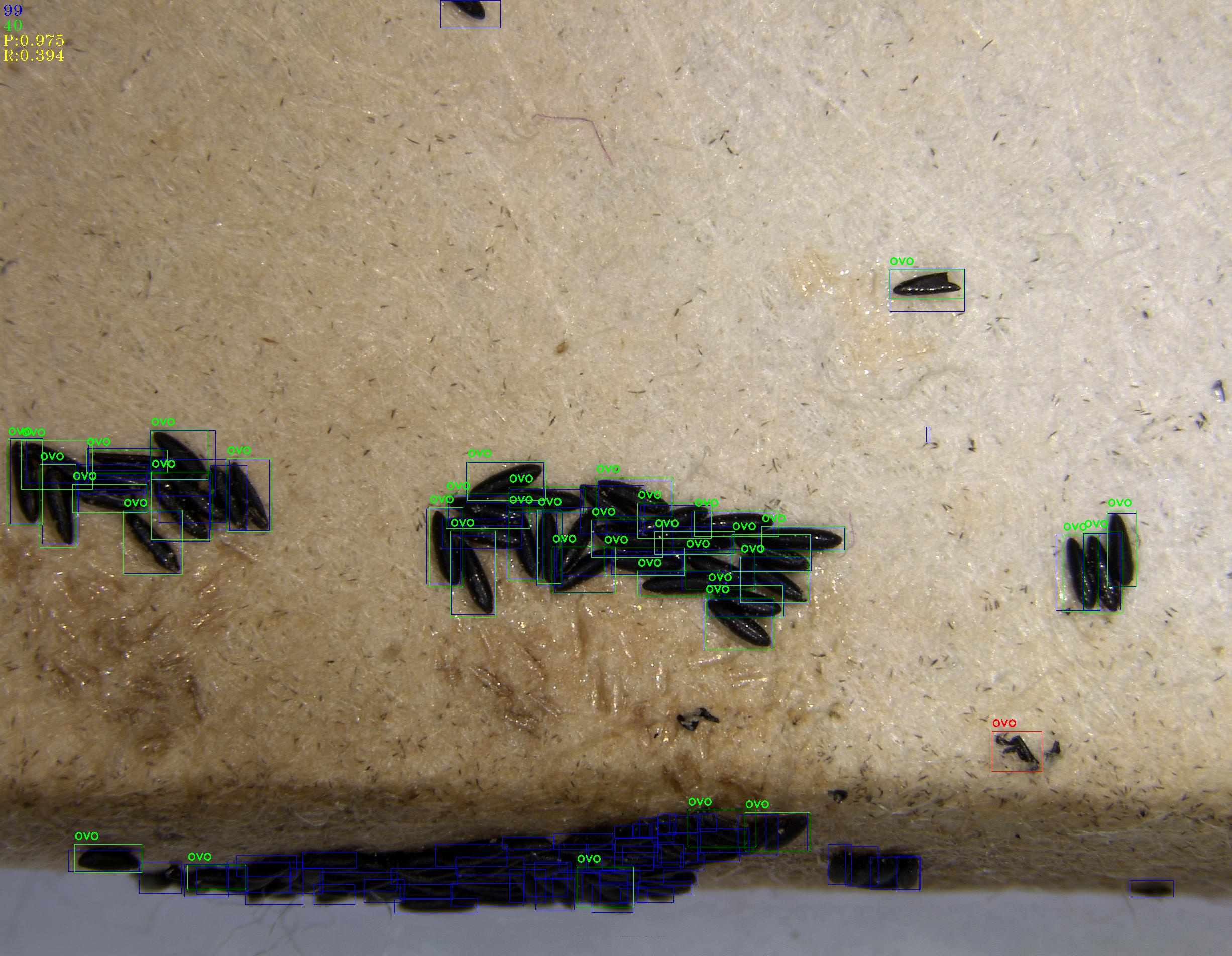}
    \caption{A pallet with eggs on the side of the pallet. One should notice that these are not only very difficult to annotate, but also that the neural network did a poor job in identifying them.}
    \label{fig:na_borda}
\end{figure}
\begin{figure}
    \centering
    \includegraphics[width=0.7\columnwidth]{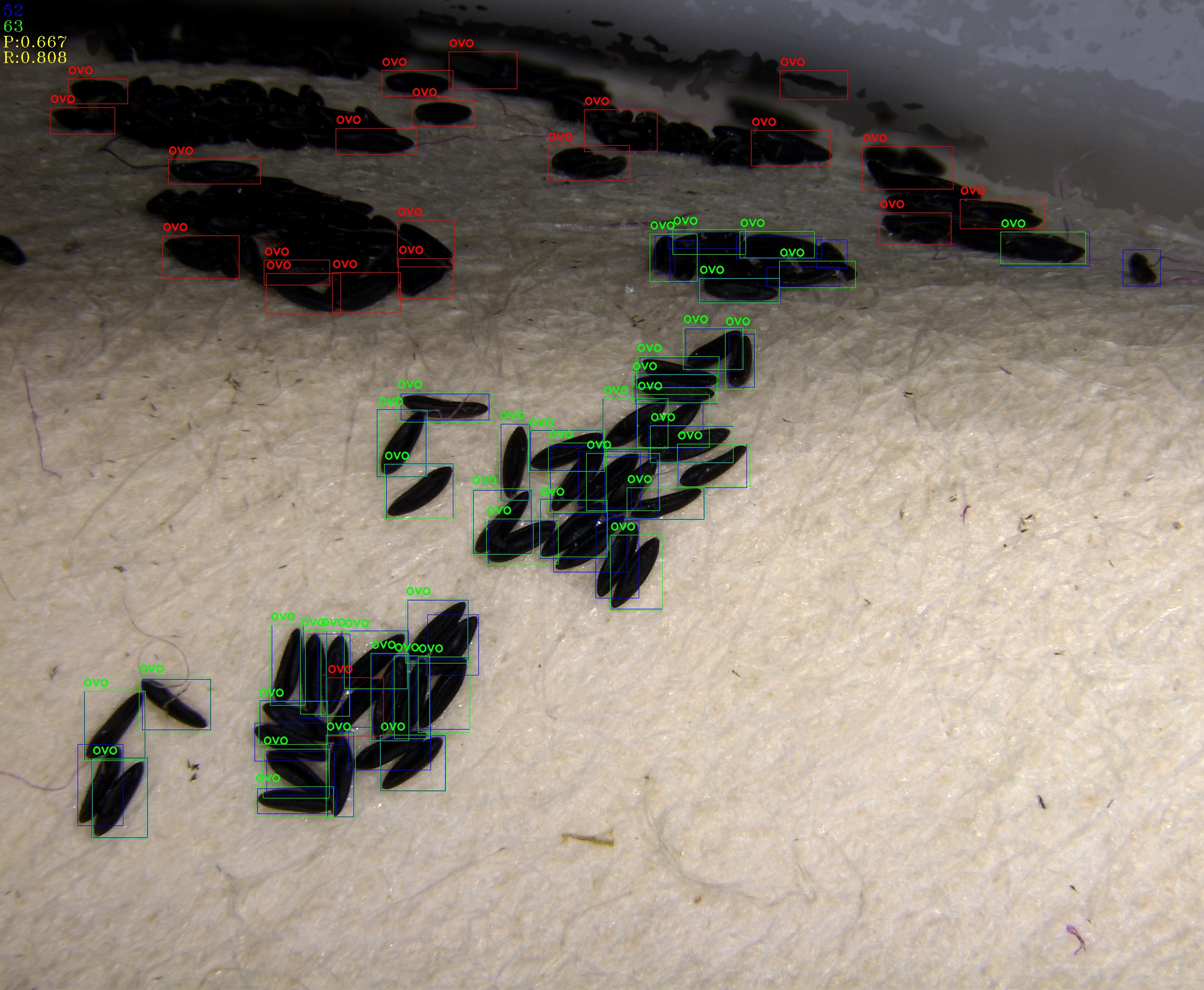}
    \caption{Eggs edged between two sides (in the center of the image). In this case, they were considered as belonging to the side that appears in the bottom of the image. One should also notice that, in this case, some eggs in the top half of the image were counted, but with subpar performance.}
    \label{fig:edged_eggs}
\end{figure}

Concerning images depicting a higher density of eggs, another dimension of the issue revolves around eggs positioned at the periphery of the pallet. This aspect exerts influence in two distinct manners: firstly, it poses a challenge for the annotation process, and secondly, it creates a complication for the performance of neural networks. The elucidation of this point is exemplified in Figure~\ref{fig:na_borda}: the intricate nature of annotating tightly clustered eggs is evident. These clusters also exacerbate the network's struggle in identifying eggs accurately. The primary complexities arise from variations in perspective, leading to shifts in the visual attributes of the eggs. Moreover, these eggs frequently suffer from being out of focus and partially obscured by their counterparts. Addressing this concern in subsequent research could require ignoring these eggs, which would necessitate image cropping. However, such a strategy is not straightforward, as it's uncertain whether these eggs won't be inadvertently detected, causing potential interference. Yet, the more challenging scenario encompasses eggs positioned at the juncture of the frontal and lateral surfaces, as depicted in Figure~\ref{fig:edged_eggs}. A facile solution, like cropping one side, isn't immediately applicable, as this could result in the partial obstruction of eggs in the process. Attempting to capture an image of the edge itself introduces another complexity, potentially including eggs from adjacent sides, akin to the instance portrayed in Figure~\ref{fig:edged_eggs}.


\begin{figure}
    \centering
\includegraphics[width=0.7\columnwidth,height=0.7\columnwidth]{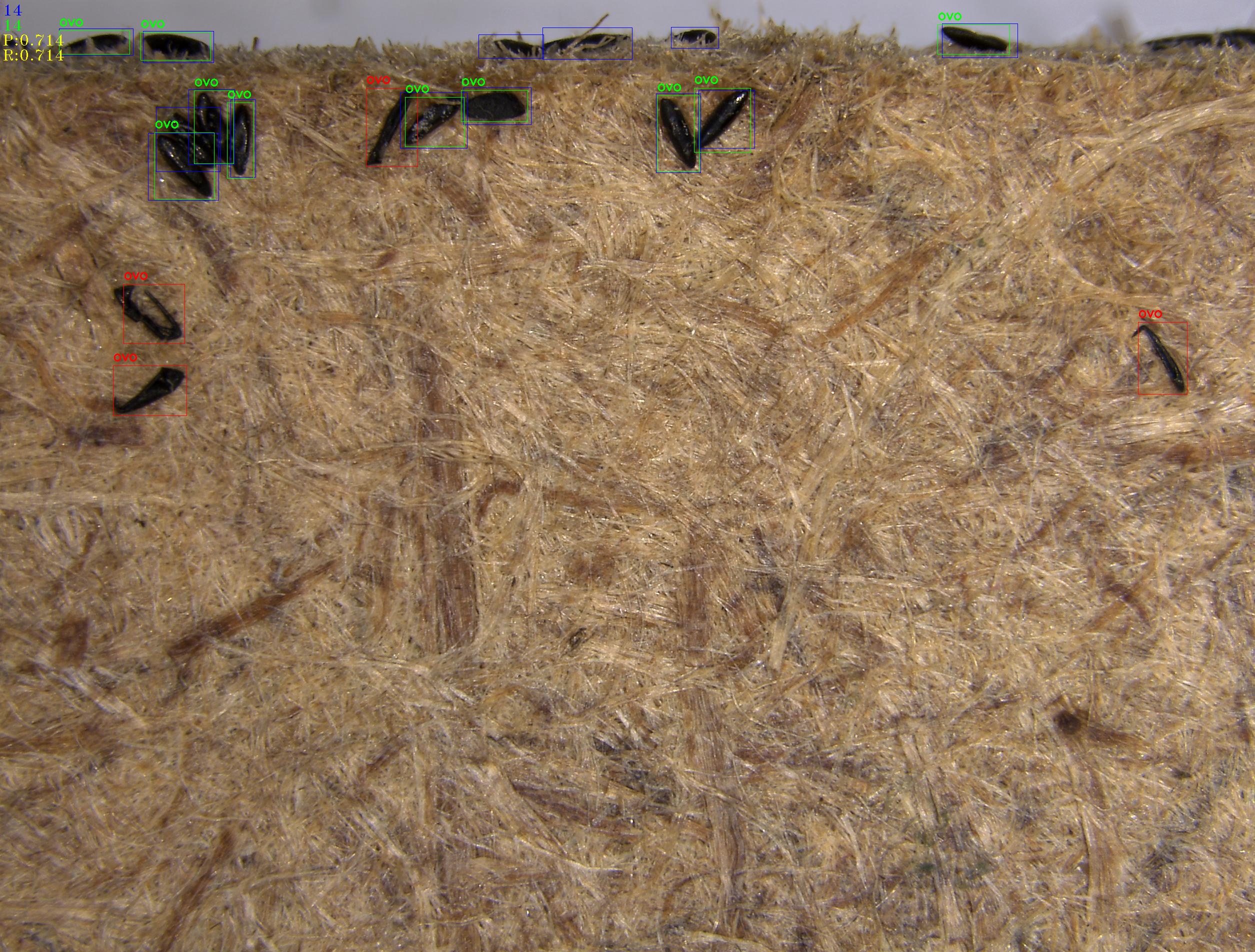}
    \caption{Dirt on the pallets counted as eggs.}
    \label{fig:sujeiras}
\end{figure}

\section{Conclusion}



To the best of the current understanding, \textit{A. aegypti} is projected to persist as a significant disease vector in the upcoming years. However, there are ways of dealing with it, in order to reduce its potential of damage. On the one hand, numerous strategies for prognosticating or forestalling disease outbreaks hinge on indices grounded in egg quantities. On the other hand, there exist critical investigation venues requiring the prolific production of eggs, thereby entailing egg quantification. The process of enumerating eggs is a labor-intensive endeavor, which can be streamlined through the application of computer vision techniques. 

Within this study, we introduced a novel image dataset that encompasses both of these contextual scenarios. Concurrently, we evaluated the efficacy of three distinct neural networks in tackling this task. The results underscore that FoveaBox stands out as the prime contender when it comes to counting extensive arrays of closely clustered eggs, surpassing both Faster R-CNN and SABL in this regard.

Furthermore, we discussed the major difficulties involved in this quantification, including the effect of high quantities of eggs and clusters, the presence of dirt and also perspective related difficulties. These major factors suggest changes in the methodology for future research, comparing the top-performing networks evaluated in this study with new networks and techniques dedicated to crowd counting, such as those presented in the works of Cheng et al.~\cite{cheng2022rethinking}, of Song et al.~\cite{song2021rethinking} and of Nguyen et al.~\cite{nguyen2022improving}. In these future analyses, if the rigor in the utilization of metrics and statistical methods is maintained, notable comparative factors may be found out, and a improvement on the \textit{A. aegypti} control may be expected.

\section*{Acknowledgments}

This work has received financial support from the Dom Bosco Catholic University and the Foundation for the Support and Development of Education, Science and Technology from the State of Mato Grosso do Sul, FUNDECT. The National Institute of Science and Technology of Hymenoptera Parasitoids INCT-HymPar provided the equipment utilized to take the pictures and the Center of Epidemiologic Control of Vectors of the Municipal Health Secretariat (CCEV/SESAU) provided the original pallets with eggs collected in field. Some of the authors have been awarded with Scholarships from the the Brazilian National Council of Technological and Scientific Development, CNPq and the Coordination for the Improvement of Higher Education Personnel, CAPES.





\begin{thebibliography}{10}
\expandafter\ifx\csname url\endcsname\relax
  \def\url#1{\texttt{#1}}\fi
\expandafter\ifx\csname urlprefix\endcsname\relax\def\urlprefix{URL }\fi
\expandafter\ifx\csname href\endcsname\relax
  \def\href#1#2{#2} \def\path#1{#1}\fi

\bibitem{siqueira_junior2022epidemiology}
J.~B. Siqueira~Junior, E.~Massad, A.~Lobao-Neto, R.~Kastner, L.~Oliver,
  E.~Gallagher, Epidemiology and costs of dengue in brazil: a systematic
  literature review, INTERNATIONAL JOURNAL OF INFECTIOUS DISEASES 122 (2022)
  521--528.
\newblock \href {https://doi.org/10.1016/j.ijid.2022.06.050}
  {\path{doi:10.1016/j.ijid.2022.06.050}}.

\bibitem{Garcia2019}
P.~S.~C. Garcia, R.~Martins, G.~L. L.~M. Coelho, G.~Camara-Chavez, Acquisition
  of digital images and identification of aedes aegypti mosquito eggs using
  classification and deep learning, Institute of Electrical and Electronics
  Engineers Inc., 2019, pp. 47--53.
\newblock \href {https://doi.org/10.1109/SIBGRAPI.2019.00015}
  {\path{doi:10.1109/SIBGRAPI.2019.00015}}.

\bibitem{sanchez_gendriz2022data_driven}
I.~Sanchez-Gendriz, G.~F. de~Souza, I.~G.~M. de~Andrade, A.~D.~D. Neto,
  A.~de~Medeiros~Tavares, D.~M.~S. Barros, A.~H.~F. de~Morais, L.~J.
  Galvão-Lima, R.~A. de~Medeiros~Valentim, Data-driven computational
  intelligence applied to dengue outbreak forecasting: a case study at the
  scale of the city of natal, rn-brazil, Scientific Reports 12~(1), cited by: 4
  (2022).
\newblock \href {https://doi.org/10.1038/s41598-022-10512-5}
  {\path{doi:10.1038/s41598-022-10512-5}}.

\bibitem{bakran_lebl2022first}
K.~Bakran-Lebl, S.~Pree, T.~Brenner, E.~Daroglou, B.~Eigner, A.~Griesbacher,
  J.~Gunczy, P.~Hufnagl, S.~Jaeger, H.~Jerrentrup, L.~Klocker, W.~Paill, J.~S.
  Petermann, B.~S. Barogh, T.~Schwerte, C.~Suchentrunk, C.~Wieser, L.~N.
  Wortha, T.~Zechmeister, D.~Zezula, K.~Zimmermann, C.~Zittra, F.~Allerberger,
  H.-P. Fuehrer, First nationwide monitoring program for the detection of
  potentially invasive mosquito species in austria, INSECTS 13~(3) (MAR 2022).
\newblock \href {https://doi.org/10.3390/insects13030276}
  {\path{doi:10.3390/insects13030276}}.

\bibitem{brisco2023field}
K.~K. Brisco, C.~M. Jacobsen, S.~Seok, X.~Wang, Y.~Lee, O.~S. Akbari, A.~J.
  Cornel, Field evaluation of in2care mosquito traps to control aedes aegypti
  and aedes albopictus (diptera: Culicidae) in hawai'i island, JOURNAL OF
  MEDICAL ENTOMOLOGY\href {https://doi.org/10.1093/jme/tjad005}
  {\path{doi:10.1093/jme/tjad005}}.

\bibitem{brun2020revisao}
A.~L. Brun, P.~L. Moraes, C.~B. Rizzi, R.~L. Rizzi, Uma revis{\~a}o das
  t{\'e}cnicas computacionais para contagem de ovos de aedes aegypti em imagens
  de ovitrampas, Revista Brasileira de Computa{\c{c}}{\~a}o Aplicada 12~(3)
  (2020) 1--15.

\bibitem{de2019solution}
C.~J. de~Santana~Junior, A.~C.~A. Firmo, R.~F. A.~P. de~Oliveira, P.~J.~B.
  Lins, G.~A. de~Lima, R.~A. de~Lima, A solution for counting aedes aegypti and
  aedes albopictus eggs in paddles from ovitraps using deep learning, IEEE
  Latin America Transactions 17~(12) (2019) 1987--1994.

\bibitem{gumiran2022aedes}
C.~R. Gumiran, A.~C. Fajardo, R.~P. Medina, M.~S. Dao, B.~E. Aguinaldo, Aedes
  aegypti egg morphological property and attribute determination based on
  computer vision, 2022, p. 581 – 585, cited by: 1.
\newblock \href {https://doi.org/10.1109/ICSIP55141.2022.9887255}
  {\path{doi:10.1109/ICSIP55141.2022.9887255}}.

\bibitem{salinas2022computer}
J.~E. Abad-Salinas, J.~A. Montero-Valverde, J.~L. Hernández-Hernández,
  V.~Cruz-Guzmán, M.~Martínez-Arroyo, E.~de~la Cruz-Gámez,
  M.~Hernández-Hernández, Computer vision-based ovitrap for dengue control,
  Communications in Computer and Information Science 1658 CCIS (2022) 123 –
  135, cited by: 0.
\newblock \href {https://doi.org/10.1007/978-3-031-19961-5_9}
  {\path{doi:10.1007/978-3-031-19961-5_9}}.

\bibitem{javed2023eggcountai}
N.~Javed, A.~López-Denman, P.~Paradkar, A.~Bhatti, Eggcountai: A convolutional
  neural network based software for counting of aedes aegypti mosquito eggs
  (2023).
\newblock \href {https://doi.org/10.21203/rs.3.rs-2963897/v1}
  {\path{doi:10.21203/rs.3.rs-2963897/v1}}.

\bibitem{iyyappan2022oviposition}
V.~Iyyappan, B.~Vetrivel, A.~C. Asharaja, S.~P. Shanthakumar, A.~D. Reegan,
  Oviposition responses of gravid aedes aegypti linn. mosquitoes (diptera:
  Culicidae) to natural organic infusions under laboratory condition, JOURNAL
  OF ASIA-PACIFIC ENTOMOLOGY 25~(1) (MAR 2022).
\newblock \href {https://doi.org/10.1016/j.aspen.2021.101853}
  {\path{doi:10.1016/j.aspen.2021.101853}}.

\bibitem{khan2023assessment}
A.~Khan, M.~Ullah, G.~Z. Khan, N.~Ahmed, A.~Shami, R.~A. E.~H. Mohamed, F.~M.
  Abd Al~Galil, M.~Salman, Assessment of various colors combined with
  insecticides in devising ovitraps as attracting and killing tools for
  mosquitoes, INSECTS 14~(1) (JAN 2023).
\newblock \href {https://doi.org/10.3390/insects14010025}
  {\path{doi:10.3390/insects14010025}}.

\bibitem{anaricci}
A.~P. Ricci, G.~F.~M. De~Freitas, M.~L. Nogueira, Estudo da viabilidade do
  ciclo de vida do aedes (stegomya) aegypti (linnaeus, 1762) (diptera,
  culicidae) em condições otimizadas, Anais do VIII Simpósio em Ciência e
  Tecnologia Ambiental e IV Encontro Multidisciplinar em Ciências Ambientais
  da Fronteira Sul, Disponível em: https://ppgctauffs.wixsite.com/simposio
  [publicação prevista do anais: 20/12/2021] (2021).

\bibitem{article_faster}
S.~Ren, K.~He, R.~B. Girshick, J.~Sun, Faster {R-CNN:} towards real-time object
  detection with region proposal networks, CoRR abs/1506.01497 (2015).
\newblock \href {http://arxiv.org/abs/1506.01497} {\path{arXiv:1506.01497}}.

\bibitem{article_rcnn}
R.~B. Girshick, J.~Donahue, T.~Darrell, J.~Malik, Rich feature hierarchies for
  accurate object detection and semantic segmentation, CoRR abs/1311.2524
  (2013).
\newblock \href {http://arxiv.org/abs/1311.2524} {\path{arXiv:1311.2524}}.

\bibitem{article_fast_rcnn}
R.~B. Girshick, Fast {R-CNN}, CoRR abs/1504.08083 (2015).
\newblock \href {http://arxiv.org/abs/1504.08083} {\path{arXiv:1504.08083}}.

\bibitem{article_sabl}
J.~Wang, W.~Zhang, Y.~Cao, K.~Chen, J.~Pang, T.~Gong, J.~Shi, C.~C. Loy,
  D.~Lin, Side-aware boundary localization for more precise object detection,
  CoRR abs/1912.04260 (2019).

\bibitem{kong2020foveabox}
T.~Kong, F.~Sun, H.~Liu, Y.~Jiang, L.~Li, J.~Shi, Foveabox: Beyound
  anchor-based object detection, IEEE Transactions on Image Processing 29
  (2020) 7389--7398.

\bibitem{cheng2022rethinking}
Z.-Q. Cheng, Q.~Dai, H.~Li, J.~Song, X.~Wu, A.~G. Hauptmann, Rethinking spatial
  invariance of convolutional networks for object counting, in: 2022 IEEE/CVF
  Conference on Computer Vision and Pattern Recognition (CVPR), 2022, pp.
  19606--19616.
\newblock \href {https://doi.org/10.1109/CVPR52688.2022.01902}
  {\path{doi:10.1109/CVPR52688.2022.01902}}.

\bibitem{song2021rethinking}
Q.~Song, C.~Wang, Z.~Jiang, Y.~Wang, Y.~Tai, C.~Wang, J.~Li, F.~Huang, Y.~Wu,
  Rethinking counting and localization in crowds: A purely point-based
  framework, in: 2021 IEEE/CVF International Conference on Computer Vision
  (ICCV), IEEE Computer Society, Los Alamitos, CA, USA, 2021, pp. 3345--3354.
\newblock \href {https://doi.org/10.1109/ICCV48922.2021.00335}
  {\path{doi:10.1109/ICCV48922.2021.00335}}.

\bibitem{nguyen2022improving}
N.~H. Tran, T.~D. Huy, S.~T.~M. Duong, P.~Nguyen, D.~H. Hung, C.~D.~T. Nguyen,
  T.~Bui, Q.~H. TRUONG, Improving local features with relevant spatial
  information by vision transformer for crowd counting, in: 33rd British
  Machine Vision Conference 2022, {BMVC} 2022, London, UK, November 21-24,
  2022, {BMVA} Press, 2022.

\end{thebibliography}

\end{document}